\documentclass[preprint,aps,ams,showkey,floatfix]{revtex4}
\usepackage{amssymb}

\usepackage{epsfig}
\usepackage{graphicx}

\begin{document}

\title{Multiscaling comparative analysis of time series and geophysical phenomena}
\author{Nicola Scafetta$^{1}$ and Bruce J. West$^{1,2}$}

\affiliation{$^{1}$ Physics Department, Duke University, Durham, NC 27708} %
\affiliation   {$^{2}$ Mathematics Division, Army Research Office,
Research Triangle Park, NC 27709. }

\begin{abstract}
Different methods are used to determine the scaling exponents
associated with a time series describing a complex dynamical
process, such as those observed in geophysical systems. Many of
these methods are based on the numerical evaluation of the variance
of a diffusion process whose step increments are generated by the
data. An alternative method focuses on the direct evaluation of the
scaling coefficient of the Shannon entropy of the same diffusion
distribution. The combined use of these methods can efficiently
distinguish between fractal Gaussian and L\'{e}vy-walk time series
and help to discern between alternative underling complex dynamics.
\end{abstract}

\date{\today}
\maketitle

The evaluation of the scaling exponents is of fundamental importance to
describe a number of complex systems \cite{2Mandelbrot,feders,otto}. The
mathematical definition of scaling is as follows \cite{goldenfeld}. The
function $\Phi (r_{1},r_{2},\ldots )$ is termed scaling invariant, if it
fulfills the property:
\begin{equation}
\Phi (r_{1},r_{2},\ldots )=\gamma ^{a}~\Phi (\gamma ^{b}r_{1},\gamma
^{c}r_{2},\ldots )~.  \label{scafun11}
\end{equation}
Thus, if we scale all coordinates $\{r_i\}$ by means of an
appropriate choice of the exponents $a,b,c....$, then we always
recover the same function. This scaling invariance is the basic
property that characterizes fractal functions
\cite{2Mandelbrot,feders,otto}. The theoretical and experimental
search for the correct scaling exponents is intimately related to
the discovery of deviations from ordinary statistical mechanics.
Fractal time series are particularly important in geophysics, as
well as in several other field of research including biophysics and
econophysics, where the phenomena of interest may present specific
self-similarity patterns on different time scales.

Two methods of analysis of time series commonly used to determine scaling
properties are autocorrelation analysis and power spectral analysis \cite
{politi}. The autocorrelation function of the fractal noise $\{\xi _{i}\}$
results in the relation
\begin{equation}
C(r)=\frac{<\xi _{i}~\xi _{i+r}>}{<\xi _{i}^{2}>}\propto r^{2H-2}~.
\label{cor111}
\end{equation}
The power spectral representation of the same scaling property reads:
\begin{equation}
S(f)=\int_{-\infty }^{\infty }C(r)~e^{-i2\pi fr}dr\propto f^{1-2H}~.
\label{cor112}
\end{equation}
It is easy to recognize the self-similarity or scaling property of the above
two equations in their power-law form. The scaling exponent $H$ was called
the \emph{Hurst exponent} by Mandelbrot \cite{2Mandelbrot} in honor of the
civil engineer Hurst who first understood the importance of scaling laws to
describe the long-range memory in time series. In particular, Hurst was
interested in evaluating the strength of the persistence of the annual level
of the floods of the Nile river and, for such a scope, developed a time
series analysis method to determine the scaling parameter $H$ \cite
{hurstbook}.

A value $H=1$ corresponds to 1/f-noise or \emph{pink} noise. The
adoption of a color name ``pink'' derives from the fact that a light
source characterized by a 1/f spectrum looks pink. These type of
noises are particular important because they represent a kind of
perfect balance between randomness and order, or between
unpredictability and predictability. In fact, for pink noises the
autocorrelation function between two events separated by a time
interval $\Delta \tau =r$ is independent on $r$, ($C(r)\approx
const$). Pink noises, $H\approx 1$, are found in countless natural
phenomena from heart-beat intervals to music \cite{paradise}. A
value $0<H<0.5$ corresponds to antipersistent noise, $H=0.5$
corresponding to uncorrelated or random noise, also known as
\emph{white} noise, and $0.5<H<1$ corresponds to correlated or
persistent noise. It is possible to extend the definition of $H$ for
values larger than 1. So, a value $H=1.5$ corresponds to Brownian
motion, which, as it is well known, describes the erratic motion of
a particle, such as a pollen grain, in suspension on  a fluid; this
erratic motion is cause by random collisions
between the particle and the molecules of the fluid \cite{Ga97}. A value $%
H=2 $ corresponds to \emph{brown} noise  and a value $H>2$ is known
as \emph{black} noise. These noises are characterized by a very
smooth shape and may be  adopted, for example, to generate
artificial mountain landscapes \cite{paradise}.

It is important to point out that there are two common alternative
deviations from ordinary statistical mechanics: anomalous Gaussian
statistics and L\'{e}vy statistics \cite{khinchin}.  These two
different statistics are indicative, in particular, of two different
kind of complex noises: the monofractal Gaussian noise \cite
{2Mandelbrot} and the L\'{e}vy-walk
 noise \cite{levywalk3}. These two types of noises
present similar long-range correlation patterns, but are generated
by quite different complex dynamics. The monofractal Gaussian noise,
in its persistent form, presents long range memory in the sense that
future events are strongly related to the frequency of occurrence of
past events and the waiting time distribution between events has
finite variance. The L\'{e}vy-walk intermittent noise, instead,
presents long-range correlation patterns which are generated by
random waiting time intervals between events obeying to an inverse
power law distribution with exponents that yield infinite variance,
and there is no real correlation between events \cite{nicola2004}.
Figs. 1 show examples of these noises.

\begin{figure}[tbp]
\epsfig{file=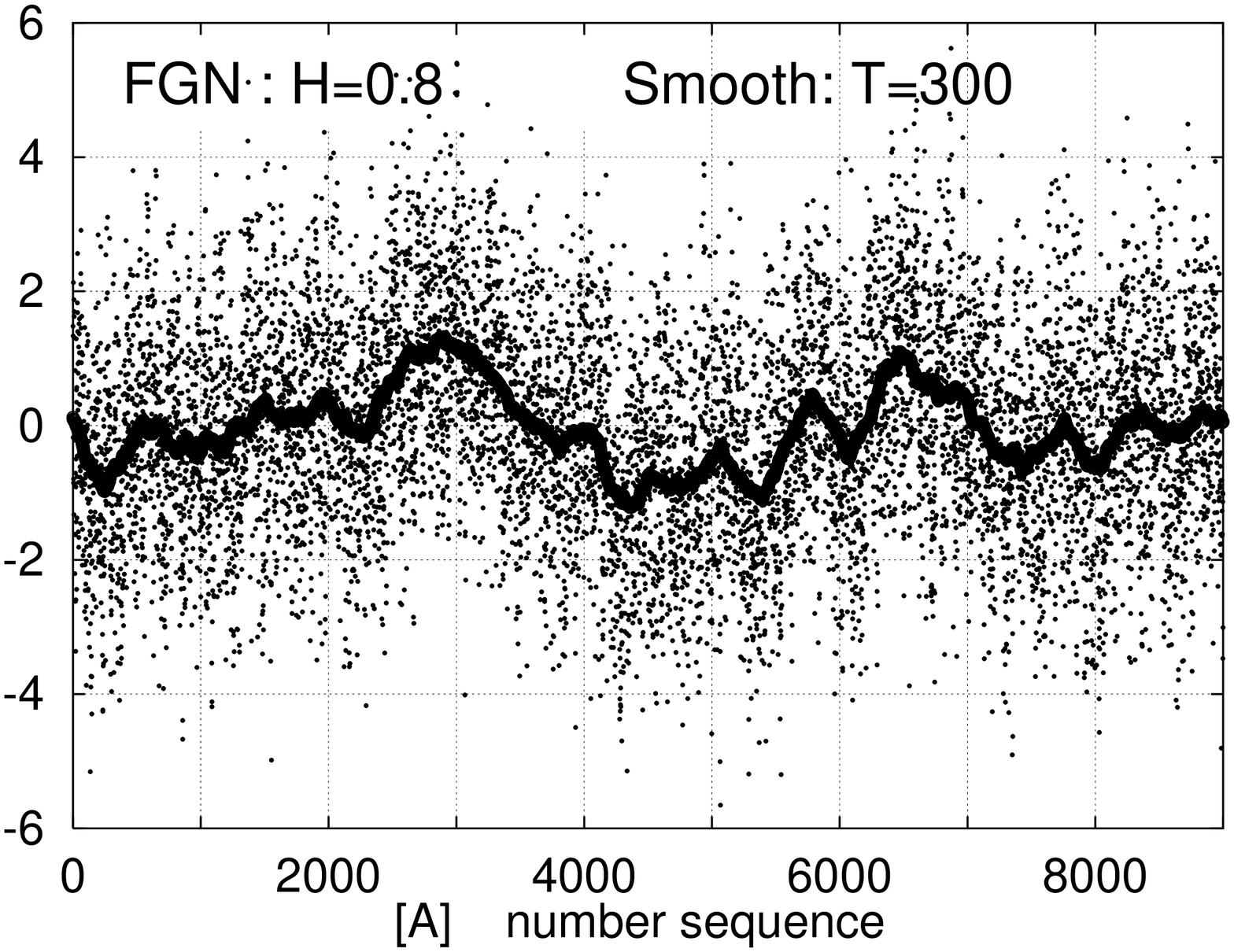,height=6.25cm,width=5.3cm,angle=-90} %
\epsfig{file=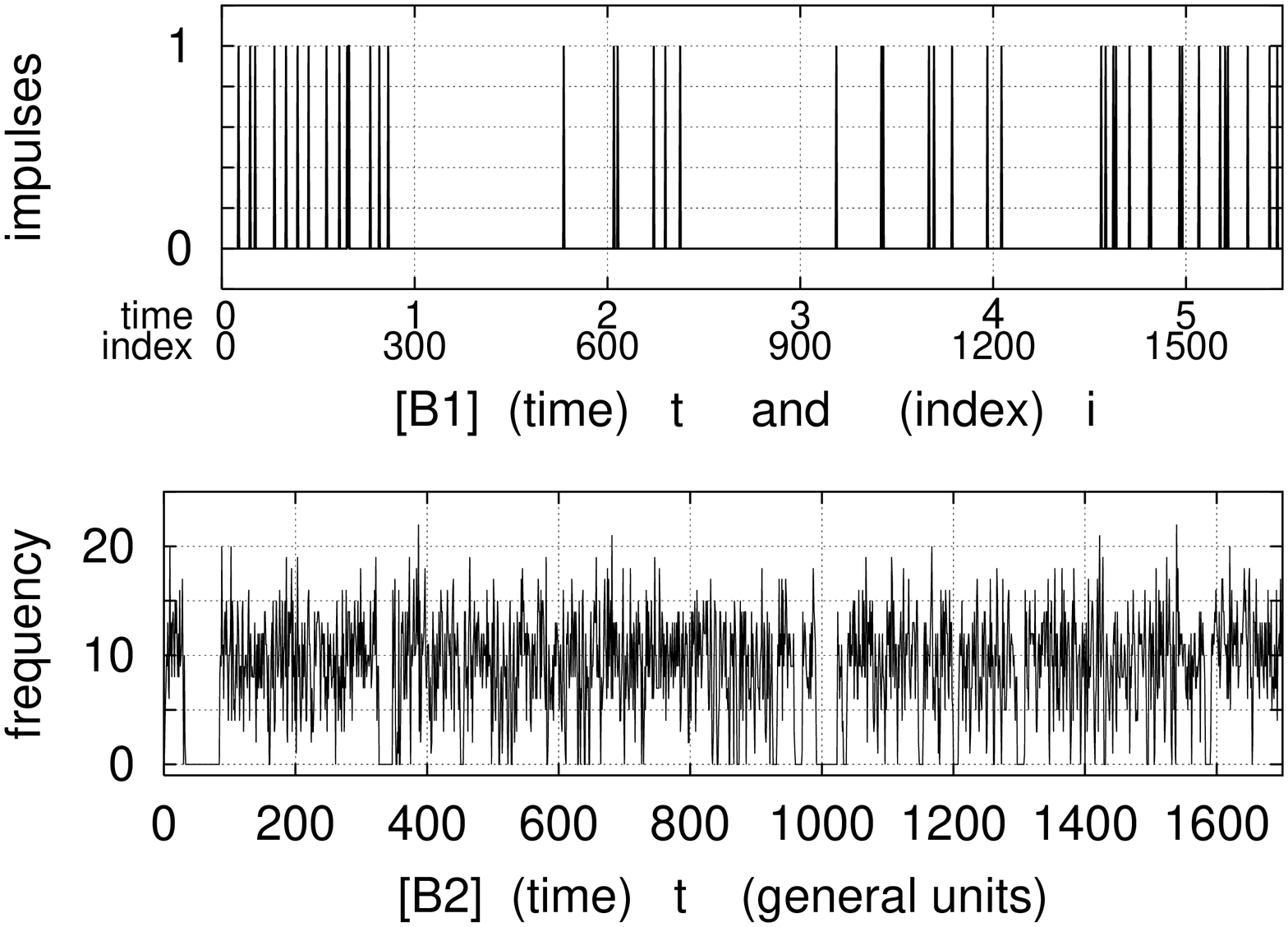,height=6.25cm,width=5.5cm,angle=-90}
\caption{ [A] Fractional Gaussian noise with $H=0.8$; [B] Two forms
of \textit{L\'evy-walk
intermittent noise} with $%
\psi(\tau)\propto \tau^{-\mu}$ and $\mu=2.5$. B2 gives the frequency
of impulses every 300 units of B1. }
\end{figure}

Herein, we briefly describe two alternative time series scaling analysis
methods, whose combined adoption can be used to distinguish the two above
alternative noises. The diffusion entropy analysis (DEA) and standard
deviation analysis (SDA) \cite{nicola2002}. Both techniques are based on the
prescription that a time series $\left\{ \xi _{i}\right\} $ of $N$ elements
are the fluctuations of a diffusion trajectory \cite{nicola2002}. Note that
there exist several other scaling analysis methods such as the detrended
fluctuation analysis \cite{dfa} and several wavelet based methods \cite
{Mallat,arneodo2,percival}, which are variance based methods and are
theoretically equivalent to the SDA.

According to the prescription of Scafetta and Grigolini \cite{nicola2002},
we shift our attention from the time series $\left\{ \xi _{i}\right\} $ to
the probability distribution function (pdf) $p(x,t)$ of the corresponding
diffusion process, that is, the pdf of the diffusion process, $p(x,t)$, is
evaluated by means of the $N-t$ sub-trajectories
\begin{equation}
x_{n}(t)=\sum_{i=0}^{t}\xi _{i+n}  \label{gfhkj}
\end{equation}
with $n=0,1,\dots $ Therefore, $x$ denotes the variable collecting the
fluctuations and is referred to as the diffusion variable. The scaling
property of the diffusion process, if it exists, takes the form
\begin{equation}
p(x,t)=\frac{1}{t^{\delta }}~F\left( \frac{x}{t^{\delta }}\right) ~,
\label{scafun12}
\end{equation}
where $\delta $ is the scaling exponent. The DEA \cite{nicola2002} is based
on the evaluation of the Shannon entropy $S(t)$ using the pdf (\ref{scafun12}%
). If the scaling condition of Eq. (\ref{scafun12}) holds true, it is easy
to prove that
\begin{equation}
S(t)=-\int p(x,t)\ln [p(x,t)]dx=A+\delta ~\ln (t)~,  \label{scafun14}
\end{equation}
where, $A$ is a constant. Numerically, we first evaluate the pdf
with histogram of size-bin equal to the standard deviation of the
data, and then use a discrete form of Eq. (\ref{scafun14}).

The SDA \cite{nicola2002} is based on the evaluation of the standard
deviation $D(t)$ of the same variable $x$, and yields
\begin{equation}
D(t)=\sqrt{\frac{\sum_{n=0}^{N-t}\left[ x_{n}(t)-\langle x;t\rangle \right]
^{2}}{N-t-1}}\propto t^{H},  \label{scafun18}
\end{equation}
where $\langle x;t\rangle =\frac{1}{N-t}\sum_{n=0}^{N-t}x_{n}(t)$ is the
mean value of $\{x_{n}(t)\}$, and $H$ is the Hurst exponent.

If the data are fractal Gaussian noise the two scaling exponents are related
to each other via the fractal Gaussian relation
\begin{equation}
H=\delta ~.  \label{FGR}
\end{equation}
If the data are generated by a L\'{e}vy-walk they are characterized by an
inverse power law waiting time distribution of the type
\begin{equation}
\psi (\tau )\propto \frac{1}{(1+\tau )^{\mu }}~,  \label{inversepowerlaw}
\end{equation}
where $2<\mu <3$, which ensures that although the first moment of $\tau $ is
finite, the second moment diverges. The scaling exponents are related to
each other via the L\'{e}vy-walk relation (LWR) \cite{nicola2004}
\begin{equation}
0.5<\delta =\frac{1}{3-2H}=\frac{1}{\mu -1}<H<1.  \label{LWR}
\end{equation}
There are several complex ways to generate a L\'evy-walk sequence,
see Ref. \cite{paolo2002,nicola2002,nicola2004,nic2002}. Some of
these noises involve mixed L\'evy-Gaussian properties. The simplest
L\'evy-walk sequence is a dichotomous signal made by a series of
zeros and ones, where $\xi=1$ represents the occurrence of an event
and $\xi=0$ represents no event. The time intervals $\{\tau_i\}$
obeying to Eq. (\ref{inversepowerlaw}) give the intervals between
events.

\begin{figure}[tbp]
\epsfig{file=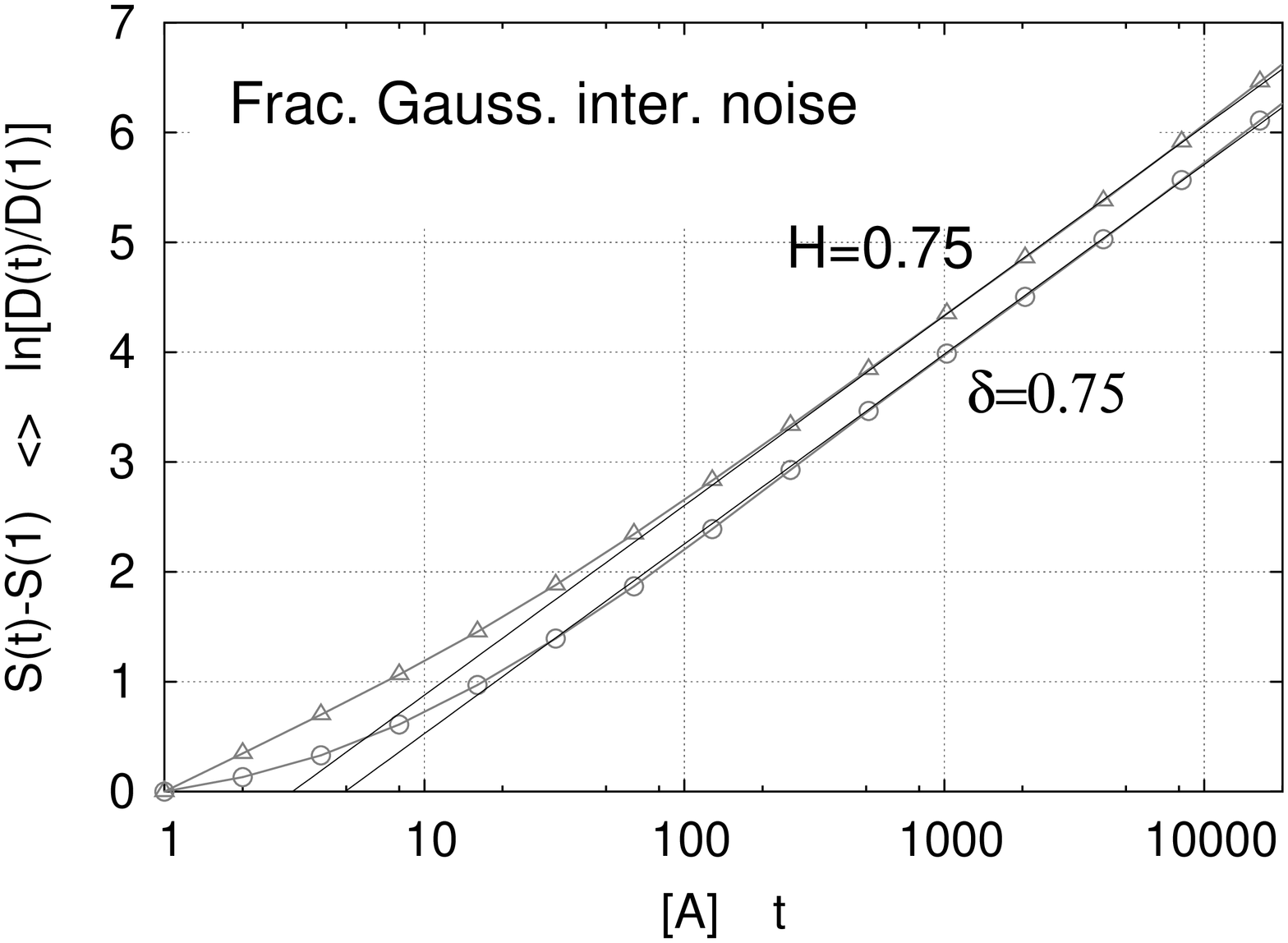,height=6.25cm,width=5.5cm,angle=-90} %
\epsfig{file=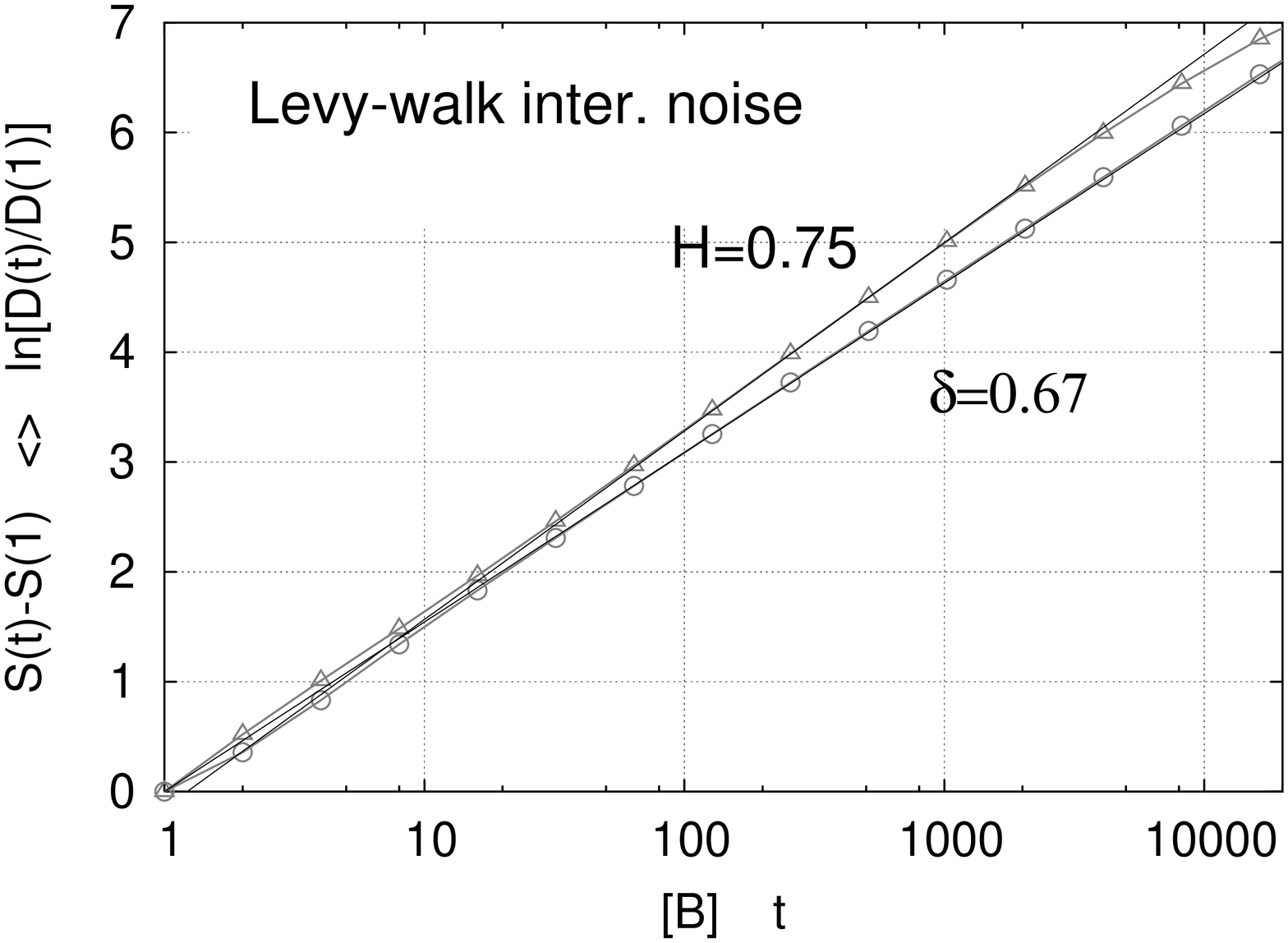,height=6.25cm,width=5.5cm,angle=-90}
\caption{ DEA end SDA of: [A] a \textit{fractal Gaussian
intermittent noise}
with $\psi(\tau) \propto \exp(- \tau/\gamma)$ with $\gamma=25$ and $%
H=\delta=0.75$; the fractal Gaussian relation (\ref{FGR}) of equal
exponents
is fulfilled; [B] a \textit{L\'evy-walk intermittent noise} with $%
\psi(\tau)\propto \tau^{-\mu}$ and $\mu=2.5$; note the bifurcation between $%
H=0.75$ and $\delta=0.67$ in accordance with  the L\'evy-walk
relation (\ref{LWR}). }
\end{figure}

Thus, by evaluating $\delta $ and $H$ and using Eq. (\ref{FGR}) and
(\ref {LWR}) it is possible to distinguish the two kinds of time
series \cite {nicola2002}, while the adoption of only one of the two
techniques can lead to a misinterpretation of the characteristics of
a phenomenon. Figs. 2A and
2B show DEA and SDA applied to a fractal Gaussian intermittent noise with $%
\delta =H=0.75$ and to a L\'{e}vy walk intermittent noise with $\mu =2.5$,
which correspond to $\delta =0.67$ and $H=0.75$, respectively \cite
{nicola2002,nicola2004}.

Particularly interesting applications of the above scaling analysis
techniques are found in geophysical phenomena such as earthquakes,
solar flares and global temperature patterns, where long time series
of data are available \cite{nicola2004,paolo2002,nicola2003}.
Herein, we briefly summarize some of our findings.

In Ref. \cite{paolo2002} it was shown that the waiting time interval
distribution $\psi(\tau)$ between solar flares \cite{SFC} is an
inverse power law of the type (\ref{inversepowerlaw}) with exponent
$\mu = 2.12 \pm 0.05$. According to the LWR (\ref{LWR}) this would
induce a Levy-walk with theoretical exponents $H=0.94 \pm 0.04$ and
$\delta=0.89 \pm 0.04$. Fig. 3A shows the waiting time distribution
of flares. Fig. 3B shows the SDA applied to several solar data such
as ACRIM TSI \cite{willson}, sunspot cover \cite {RGO} , sunspot
number \cite{SIDAC}, global surface temperature anomalies
(1856-2003) \cite{CRU} and the theoretical prediction derived from
the solar flare intermittency. The curves are quite parallel and
suggest that a L\'evy like process regulate the dynamics of the
solar activity and that the Earth climate seems to contain the same
statistics. The latter statement seems further confirmed by Figs. 4A
and 4B that show DEA and SDA applied to the ACRIM TSI and the global
temperature record showing the typical LWR bifurcation. We observe
that if these findings are not accidental, they might imply the
existence of a non-negligible complex Sun-Climate nonlinear coupling
on a short time-scale as some studies seem to confirm \cite
{aleo,Jackman}.

\begin{figure}[tbp]
\epsfig{file=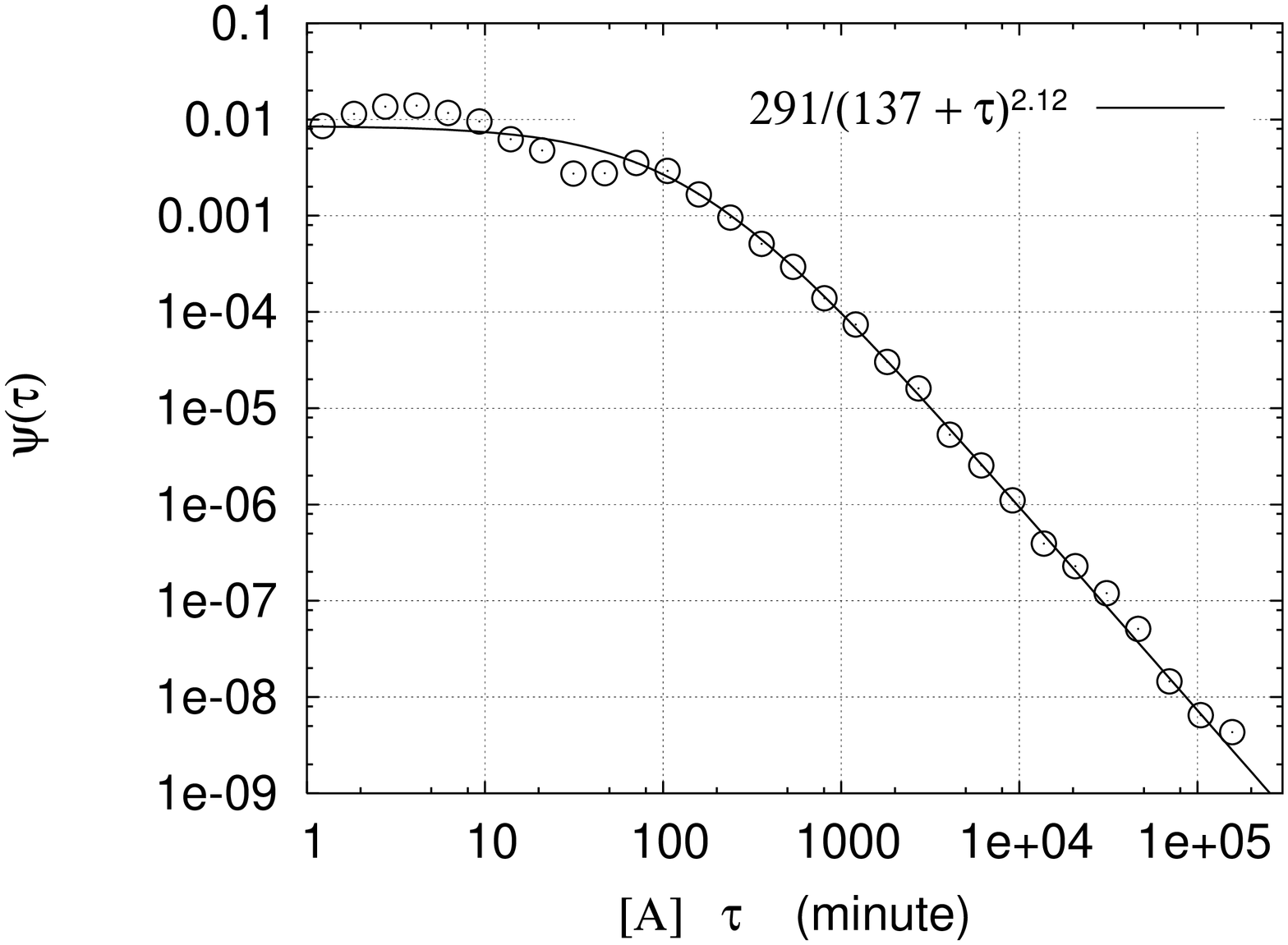,height=6.25cm,width=5.5cm,angle=-90} %
\epsfig{file=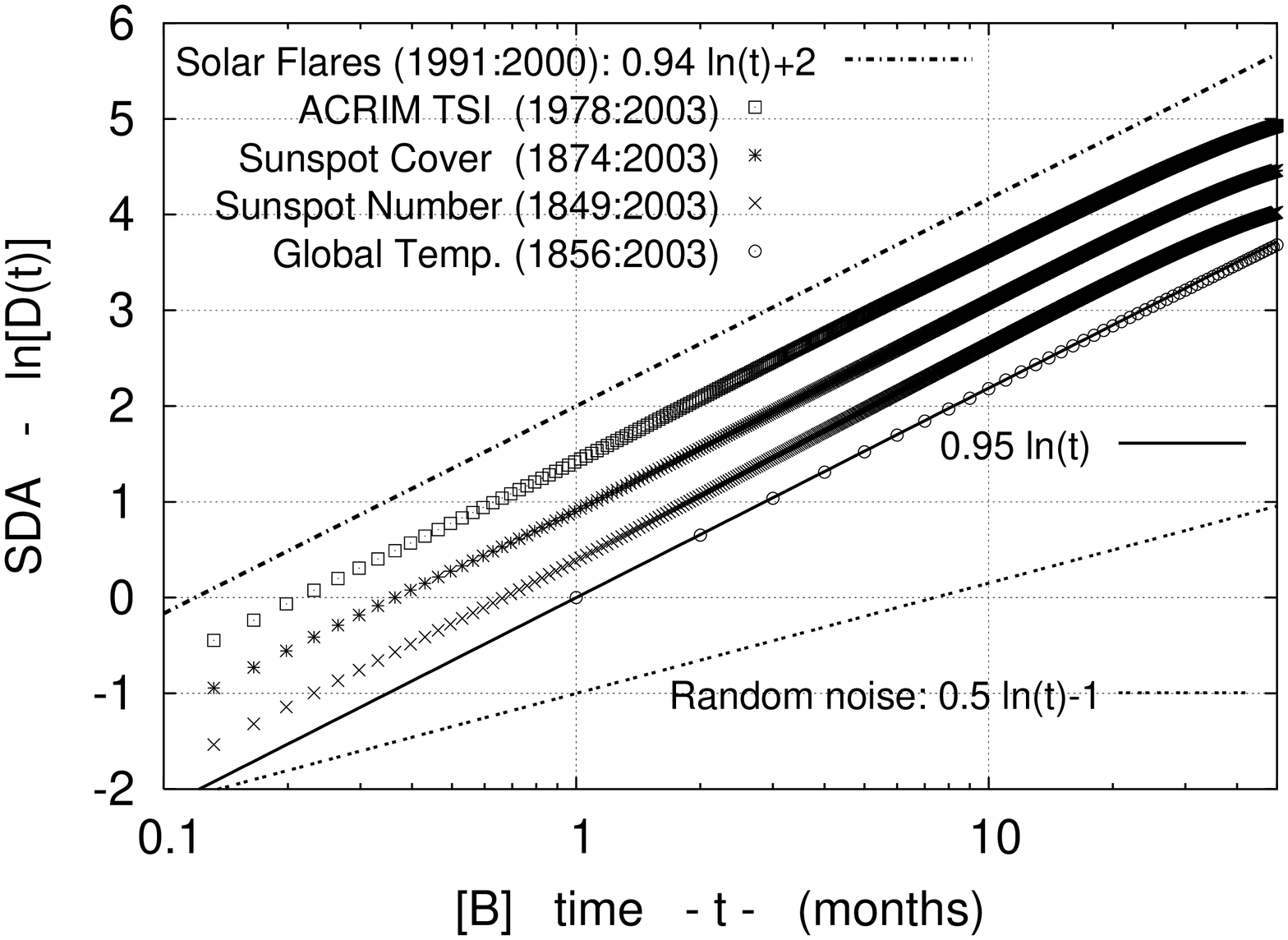,height=6.25cm,width=5.5cm,angle=-90}
\caption{ [A] Waiting time distribution of flares. The distribution
is fit
with an inverse power law Eq. (\ref{inversepowerlaw}) with exponent $%
\mu=2.12\pm0.05$ [Grigolini \textit{et al.} 2002]. [B] SDA applied to the
ACRIM composite TSI time series, sunspot number and cover sequence and
global temperature anomalies. The uppermost line represents the theoretical
L\'evy-walk scaling of the solar flare intermittency, $H_T=0.94\pm 0.02$,
obtained via Eq. (\ref{LWR}) with $\mu=2.12\pm0.05$. The bottom line shows
the scaling for random noise, $H=0.5$, for comparison. }
\end{figure}

\begin{figure}[tbp]
\epsfig{file=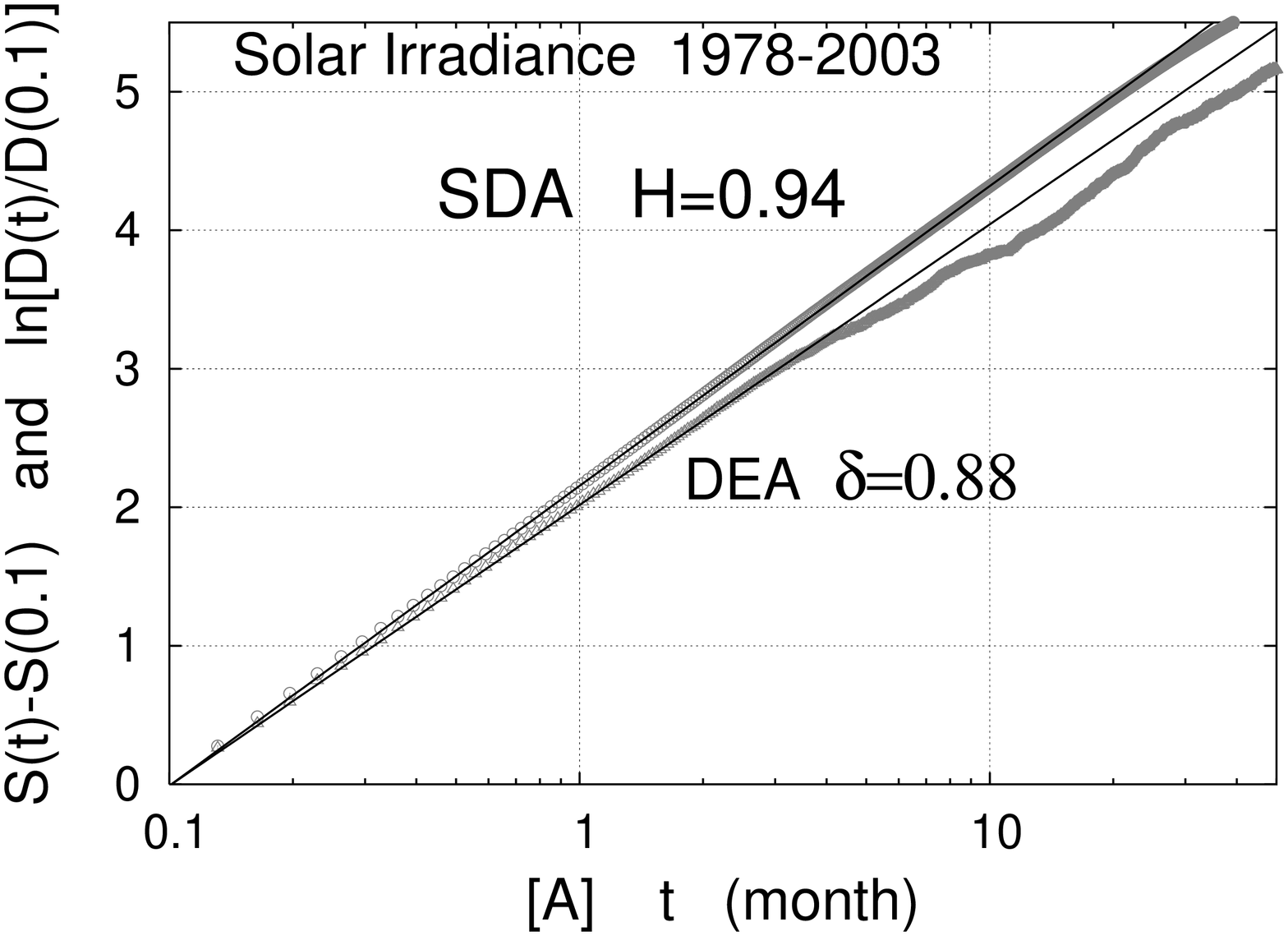,height=6.25cm,width=5.5cm,angle=-90} %
\epsfig{file=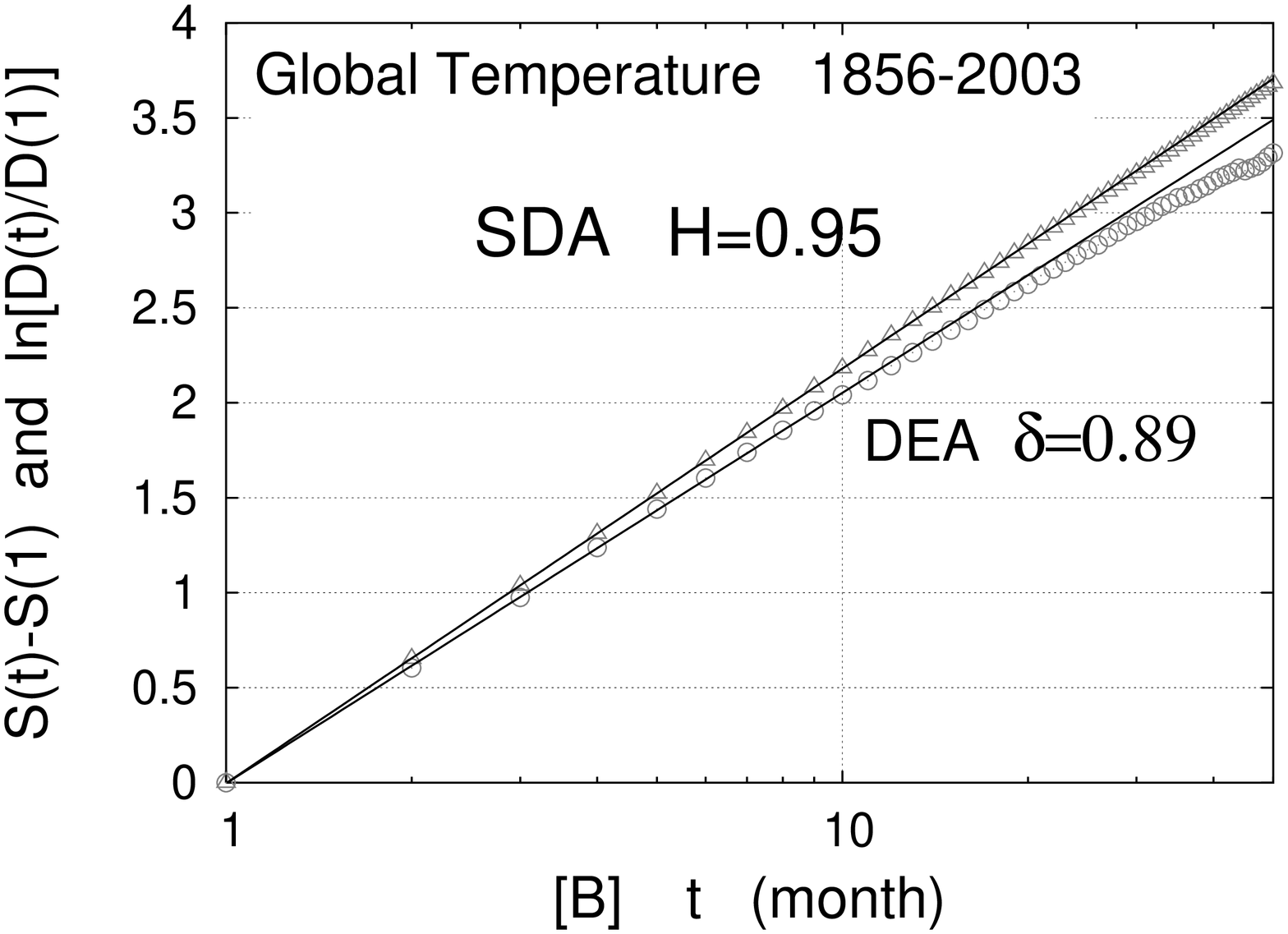,height=6.25cm,width=5.5cm,angle=-90}
\caption{ [A] DEA and SDA applied to the ACRIM composite TSI time
series. The two straight lines correspond to the scaling
coefficients $\delta =0.88 \pm 0.02$ and $H=0.94\pm 0.02$. [B] DEA
and SDA applied to the global temperature anomalies (1856-2003) time
series. The two straight lines
correspond to the scaling coefficients $\delta =0.89 \pm 0.02$ and $%
H=0.95\pm 0.02$. }
\end{figure}

The above techniques can be applied also to earthquake concurrence
\cite {nicola2004}. An issue about seismic phenomena is whether: (1)
they obey a statistics according to which the waiting times between
Omori's earthquake clusters \cite{omori} are uncorrelated from one
another, as the traditional Generalized Poisson model
\cite{stain,vito}, the ``ETAS'' model \cite{Kagan}; (2) the Omori's
earthquake clusters obey to some knind of L\'{e}vy-walk statistics
\cite{vito}; (3) or whether the data may also be characterized by
intercluster 1/f long-range correlations  between Omori's clusters
that may disclose the \textit{earthquake conversations} recently
suggested by Stein \cite{stain}. Understanding the nature of the
long-range correlation is fundamental for building reliable
earthquake models.

Fig. 5A shows the waiting time PDFs between earthquakes in
California \cite {web} using four earthquake magnitude thresholds
$M_{t}=$ 1, 2, 3 and 4. The PDFs show an initial Omori's law
\cite{omori} ($P(\tau )\propto 1/\tau $), but the pdf tails present
a large inverse power law exponent $\mu >4$ and may even approach an
exponential (or Poisson) distribution asymptotically. The Omori's
law is determined by the short-range correlated aftershocks \cite
{omori} and lasts for a time that increases with the magnitude
threshold. Fig. 5B shows the DEA and SDA applied to the intermittent
time signal $\xi (t)$, where $t$ is the physical time, obtained by
assigning a value equal to 1 at the occurrence of an event, and a
value equal to 0 when no event occurred. The latter figure suggests
that the data fulfill FGR (\ref{FGR}). Thus, beyond the Omori's law,
the earthquake clusters might be uncorrelated if the observed
super-diffusion $\delta =H=0.94$ is generated by a long-tailed
Omori's law involving multiple clusters, or there might be the
possibility that the clusters are correlated as a 1/f Gaussian noise
\cite {nicola2004}. In the latter case, traditional earthquake
models such as the Generalized Poisson model \cite{stain,vito} or
the ``ETAS'' model \cite{Kagan} should be improved by adding
additional  correlations between clusters.

\begin{figure}[tbp]
\epsfig{file=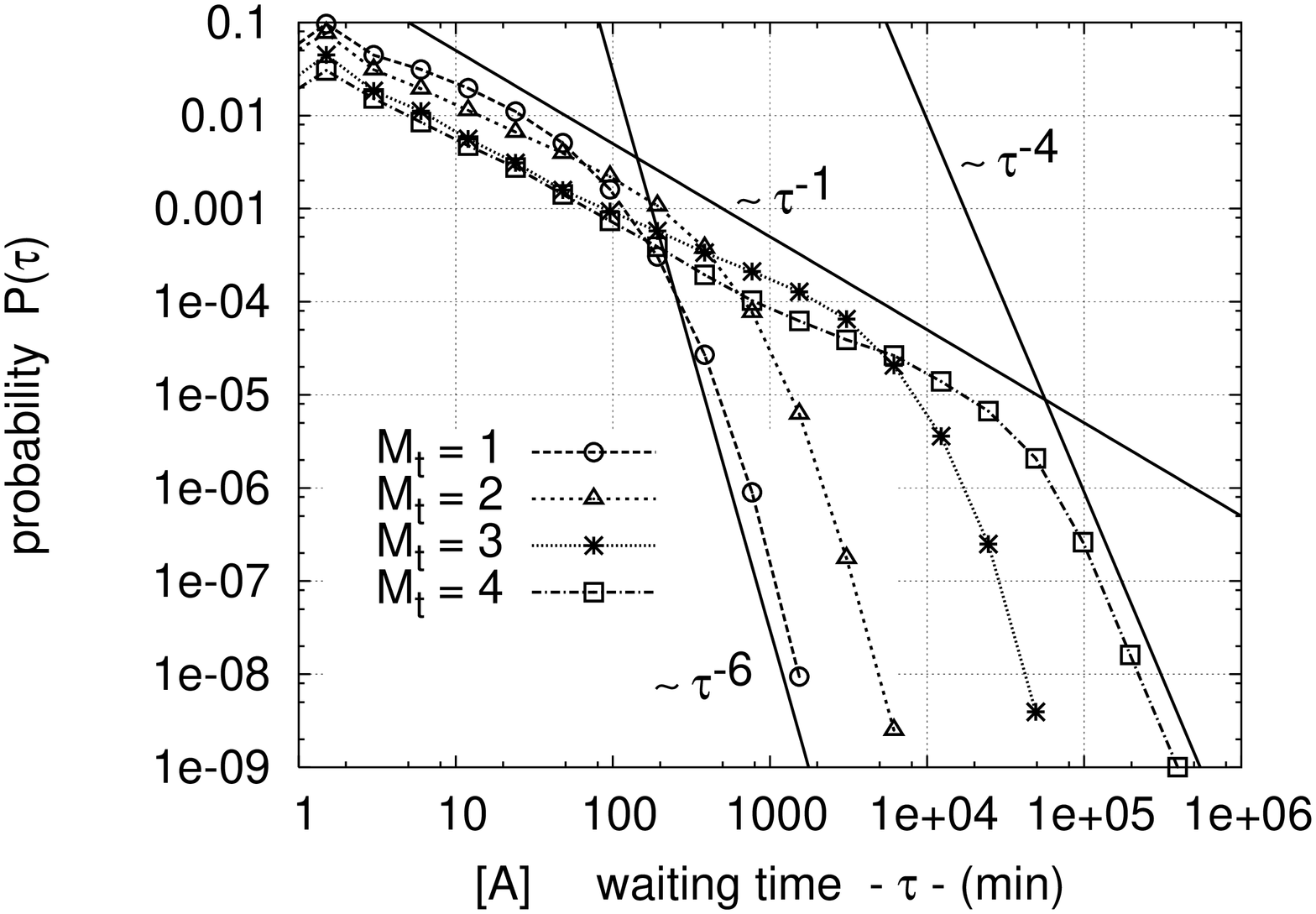,height=6.25cm,width=5.5cm,angle=-90} %
\epsfig{file=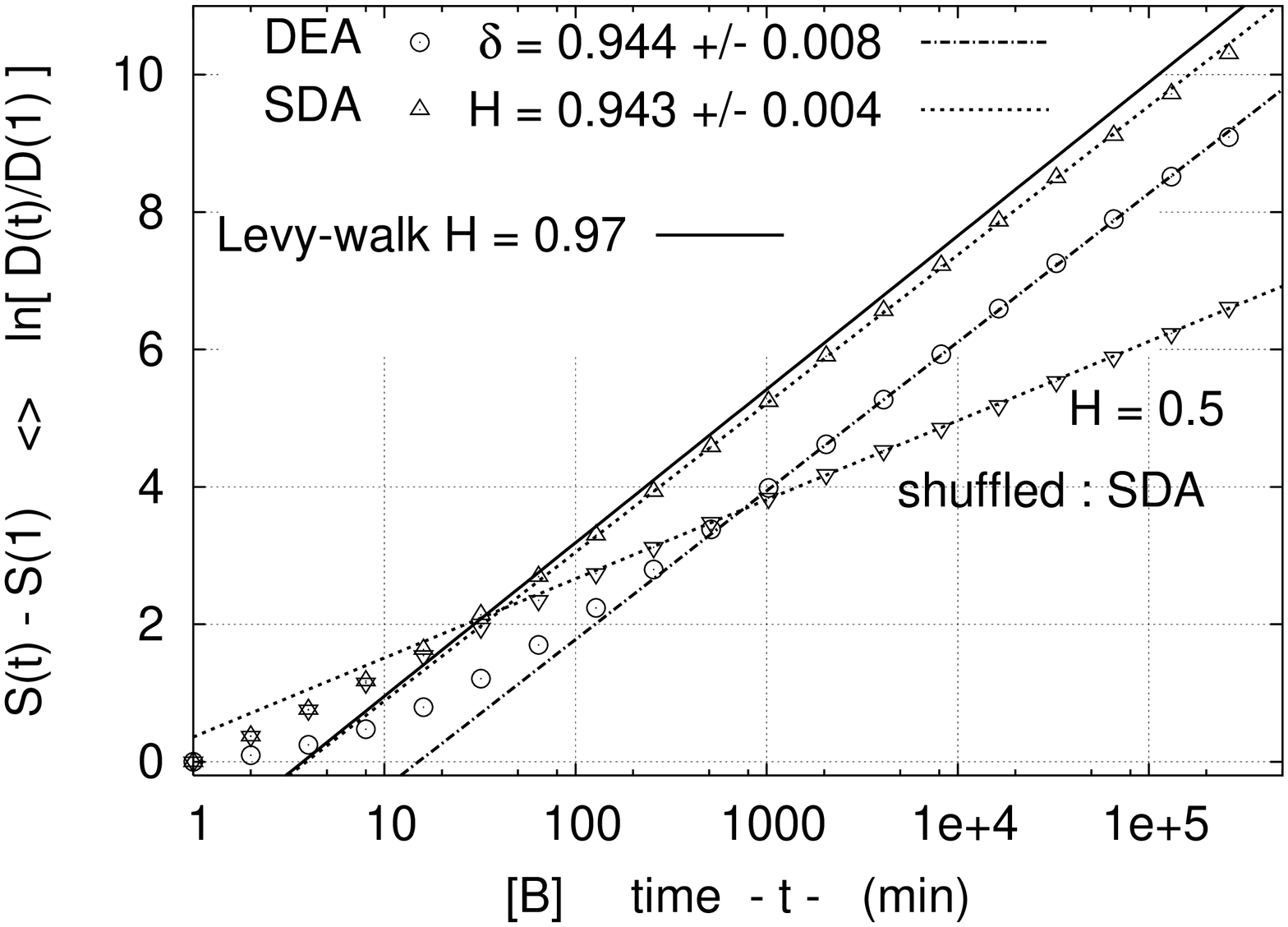,height=6.25cm,width=5.5cm,angle=-90}
\caption{ [A] pdf of the waiting times $\tau_i$ of earthquakes with
a magnitude $M\geq M_t$ = 1, 2, 3 and 4. The initial $P(\tau)\propto
1/\tau$ is the Omori's law \protect\cite{omori}. [B] DEA and SDA of
the intermittent time signal $\xi(t)$ for magnitude $M\geq M_t$ = 1.
The data are fitted with scaling exponents $\delta=0.944 \pm 0.008$
and $H=0.943\pm0.004$. The uppermost solid line with $H=0.97$
corresponds to the expectation of H if the Levy-walk condition
(\ref{LWR}) holds true. }
\end{figure}

In conclusion, we have discussed some properties of complex time
series analysis showing two different types of anomalous statistics:
fractal Gaussian noise and L\'evy-walk noise. We have shown how the
multiscaling comparative analysis of time series can be used to
distinguish the two types of noises and applied it to study some
complex patterns of geophysical phenomena. Thus, we conclude that
there are some difficulties in interpreting intermittent sequences.
Models with alternative statistics can reproduce some patterns of a
time series  equally well. This fact suggests the need of an
analysis involving complementary tests for addressing complex
systems.
\newline

\textbf{Acknowledgment:} N.S. thanks the ARO for support under grant
DAAG5598D0002.



\begin{references}

\bibitem{2Mandelbrot}B.B. Mandelbrot,
{\it The Fractal Geometry of Nature}, Freeman, New York, (1983).

\bibitem{feders}  Feders J., \emph{Fractals}, Plenum Publishers,
New York, (1988).

\bibitem{otto} Peitgen H.-O., J. Hartmut, D. Saupe, \emph{Chaos and
Fractals, new frontiers of science}, sec. edition, Springer, New
York, (2004).

\bibitem{goldenfeld} Goldenfeld N., \emph{Lectures on Phase
Transitions and the Renormalization Group} ( Perseus Book, Reading,
Massachusetts,1985).

\bibitem{politi} R. Badii and A. Politi,
{\it Complexity, Hierarchical structures and scaling in physics},
 Cambrige University Press, UK, (1997).


\bibitem{hurstbook} Hurst H.E., R. P. Black, Y.M. Simaika,
{\it LongTerm Storage: An Experimental Study}, Constable, London,
(1965).

\bibitem{paradise} Schroeder M.,\emph{Fractals, Chaos, Power Laws:
Minutes from an Infinite Paradise}, W.H. Freeman \& Company,(1992).


\bibitem{Ga97}  Gardiner C.W., \emph{Handbook of Stochastic Methods
for Physics, Chemistry and the Natural Sciences}, 2$^{nd}$ edition,
Springer-Verlag, New York, New York, (1997).

\bibitem{khinchin} A.I. Khinchin, \emph{Mathematical Foundations of
Statistical Mechanics}, Dover Publications, Inc. New York, (1949).

\bibitem{levywalk3}  Shlesinger M.F., B.J. West, J. Klafter, ``L\'evy dynamics of
enhanced diffusion: Application to turbulence,'' Phys. Rev. Lett.
{\bf 58}, 1100 (1987).

\bibitem{nicola2004} Scafetta N., and B.J. West,
``Multi-scaling comparative analysis of time series and a discussion
on 'earthquake conversations' in California,''  Phys. Rev. Lett.
\textbf{92} 138501 (2004).


\bibitem{nicola2002} Scafetta N., and P. Grigolini, ``Scaling detection in time
series: diffusion entropy analysis,''   Phys. Rev. E \textbf{66},
036130, (2002).

\bibitem{dfa}  Peng C.-K., S.V. Buldyrev, S. Havlin, M. Simons,
H.E. Stanley, and A.L. Goldberger, ``Mosaic organization of DNA
nucleotides,'' Phys. Rev, E {\bf 49}, 1685, (1994).

\bibitem{Mallat}   Mallat S.G., \textit{A Wavelet Tour of Signal Processing}
(2nd edition), Academic Press, Cambridge (1999).

\bibitem{arneodo2}  Muzy J.F.,  E. Bacry, A. Arneodo, ``The Multifractal
Formalism Revisited with Wavelets,'' Int. J.  Bifurc. Chaos
\textbf{4}, No. 2, 245-302 (1994).

\bibitem{percival} Percival  D.B., and A.T. Walden,
{\it Wavelet Methods for Time Series Analysis}, Cambrige University
Press, Cambrige (2000).


\bibitem{paolo2002} Grigolini P., D. Leddon, N. Scafetta, "The Diffusion entropy and
waiting time statistics of hard x-ray solar flares,"  Phys. Rev. E
\textbf{65}, 046203 (2002).

\bibitem{nicola2003} Scafetta N., and B.J. West,  "Solar Flare Intermittency and
the Earth's Temperature Anomalies,"       Phys. Rev. Lett.
\textbf{90}, 248701 (2003).

\bibitem{nic2002}Scafetta N., V. Latora and P. Grigolini, ``Scaling without detrending: the
diffusion entropy method applied to the DNA sequences,''   Phys.
Rev. E \textbf{66}, 031906 (2002).

\bibitem{SFC}  Solar Flare Catalog, (2003).
\textit{http://umbra.nascom.nasa.gov}

\bibitem{willson}  Willson R.C., and A.V. Mordvinov , ``Secular total solar
irradiance trend during solar cycles 21-23,'' Geophys. Res. Lett.,
\textbf{30}, 1199, doi: 10.1029/2002GL016038 (2003).
\textit{http://www.acrim.com}

\bibitem{SIDAC}  Solar Influences Data analysis Center (2003). \textit{%
http://sidc.oma.be/index.php3}

\bibitem{RGO}    Royal Greenwich Observatory/USAF/NOAA Sunspot Record
1874-2003 (2003).
\textit{http://science.msfc.nasa.gov/ssl/pad/solar/greenwch.htm}

\bibitem{CRU}  Climatic Research Unit, UK, (2003) \textit{%
http://www.cru.uea.ac.uk}.

\bibitem{aleo} D'Aleo Joe, ``Solar influence may be evident in three of the
last four
 seasons,'' (2002). \emph{http://www.intellicast.com/DrDewpoint/Library/1339}

\bibitem{Jackman} Jackman C.H, and R.D. McPeters, ``The effect of solar proton
events on ozone and other constituents,''  Chapter in \emph{Solar
Variability and its
 Effects on Climate} by Pap J.M, and P. Fox, Geophysical Monograph Series Volume 141 (2004).

\bibitem{stain}  Stein R.S., ``Earthquake Conversation,''
 Scientific American, \textbf{288} Jan 72
(2003).

\bibitem{vito}   Mega M.S., P. Allegrini, P. Grigolini, V. Latora, L.
Palatella, A. Rapisarda and S. Vinciguerra, ``Power-Law Time
Distribution of Large Earthuakes,'' Phys. Rev. Lett. \textbf{90}
188501 (2003).


\bibitem{omori}   Omori F., ``On the aftershocks of earthquakes,'' J. College Sci.
Imper. Univ. Tokyo \textbf{7}, 111
(1895). P. Bak, K. Christensen, L. Danon and T. Scanlon, ``Unified
Scaling Law for Earthquakes,'' Phys. Rev. Lett. \textbf{88} 178501
(2002).

\bibitem {Kagan}  Kagan Y.Y.,  and L. Knopoff, ``Statistical short-term earthquake
prediction,'' Science {\bf 236}, 1563 (1987).

\bibitem{web}  Southern California Earthquake Data Center (2003).
http://www.scecdc.scec.org/ftp/catalogs/SCNS/

\end{references}
\end{document}